\documentclass[print,twocolumn, showpacs, aps]{revtex4}
\usepackage{graphicx}
\usepackage{epsfig}
\usepackage{CJK}
\usepackage{amsmath}
\usepackage{supertabular}
\usepackage{textcomp}
\usepackage{threeparttable}
\usepackage{color}    %\color{red/blue/green/black/white/cyan/magenta/yellow}
\usepackage{slashed}
\usepackage{ulem}     %\uline \uuline \uwave \sout
\usepackage{cancel}
\usepackage{underscore}
\usepackage{hyperref}
\hypersetup{bookmarksnumbered,%
	colorlinks,%
	linkcolor=blue,%
	citecolor=blue,%
	plainpages=false,%
	pdfstartview=FitH}

\begin{document}
\normalem

%\renewcommand\figurename{Fig}
%$\footnote{wanniu\underline{ }nju163.com}$
%$\footnote{takayuki.myo@oit.ac.jp}$
%$\footnote{cxu@nju.edu.cn}$
%$\footnote{toki@rcnp.osaka-u.ac.jp}$
%$\footnote{horiuchi@rcnp.osaka-u.ac.jp}$
%$\footnote{mengjiao@rcnp.osaka-u.ac.jp}$

\title{Finite particle-number description of symmetric nuclear matter with spin excitations of high-momentum pairs induced by tensor force %Spin excitations for high-momentum nucleon pairs in symmetric nuclear matter
}
\author{Niu Wan,$^{1}$$\footnote{wanniu@scut.edu.cn}$ Takayuki Myo,$^{2,3}$$\footnote{takayuki.myo@oit.ac.jp}$ Hiroki Takemoto,$^{4}$ Hiroshi Toki,$^{3}$ Chang Xu,$^{5}$$\footnote{cxu@nju.edu.cn}$ Hisashi Horiuchi,$^{3}$ Masahiro Isaka,$^{6}$ Mengjiao Lyu,$^{7,8}$ and Qing Zhao$^{9}$}
\address{$^{1}$School of Physics and Optoelectronics, South China University of Technology, Guangzhou 510641, China\\
$^{2}$General Education, Faculty of Engineering, Osaka Institute of Technology, Osaka, Osaka 535-8585, Japan\\
$^{3}$Research Center for Nuclear Physics (RCNP), Osaka University, Ibaraki, Osaka 567-0047, Japan\\
$^{4}$Faculty of Pharmacy, Osaka Medical and Pharmaceutical University, Takatsuki, Osaka 569-1094, Japan\\
$^{5}$School of Physics, Nanjing University, Nanjing 210093, China\\
$^{6}$Science Research Center, Hosei University, 2-17-1 Fujimi, Chiyoda-ku, Tokyo 102-8160, Japan\\
$^{7}$College of Physics, Nanjing University of Aeronautics and Astronautics, Nanjing 210016, China\\
$^{8}$Key Laboratory of Aerospace Information Materials and Physics (NUAA), MIIT, Nanjing 211106, China\\
$^{9}$School of Science, Huzhou University, Huzhou 313000, Zhejiang, China}

\begin{abstract}
We study the symmetric nuclear matter using bare nucleon-nucleon ($NN$) interactions with finite particle-number approach within finite cubic boxes. Due to the $NN$ correlations originating from bare $NN$ interaction, two nucleons can be excited to the high-momentum region, leading to the increase of the kinetic energy in nuclear matter. We further consider the spin excitations in the nucleon pairs, where the spin of the two nucleons are changed, and this excitation is important for the tensor correlation. The unitary correlation operator method (UCOM) is used to treat the short-range correlation. The tail correction coming from the neighbouring boxes is also included. We demonstrate the contributions of various excitations of nucleon pairs as well as the tail correction to the total energy at the normal density. We also discuss the effects of UCOM and correlated nucleon pairs on the density dependence of the total energy. We calculate the equations of state of symmetric nuclear matter using two kinds of the Argonne potentials and the results agree with those from other many-body theories. The density dependences of the Hamiltonian components are also shown.
\end{abstract}

% nuclear matter
% Binding energies and masses
% Nucleon–nucleon interactions

\maketitle

\section{Introduction}
\label{intro}
The equation of state (EoS) of nuclear matter is of great importance for both nuclear physics and astrophysics \cite{Akmal,JML,CJHo,Miller,AWS,BALi,SGan}. The properties of nuclear matter have close correlations to the structure of finite nuclei and the dynamics and evolution of neutron stars. Due to the great efforts of the community, different models have been proposed to study the properties of finite nuclei, however, it is difficult from the terrestrial nuclei to extrapolate the information for neutron stars under extreme conditions such as very large density. One possible way to understand the astrophysical phenomena is to study the nuclear matter within many-body approaches based on bare nucleon-nucleon ($NN$) interactions. The fundamental $NN$ interaction has a strong repulsive force at short-range distances and a strong tensor force at intermediate- and long-range distances \cite{AV1,AV2,AV3}. They can respectively induce short-range correlation and tensor correlation, which produce high-momentum components in both finite nuclei and nuclear matter \cite{HM0,HM1,HM2,HM3,HM4,HM5,HM6,HM7}.

Several popular approaches have been proposed to treat the complicated $NN$ correlations originating from the $NN$ interaction. One is the utilization of correlation functions multiplied to the trial wave function, such as the Jastrow factor introduced in the Green's function Monte Carlo (GFMC) \cite{GFMC0,GFMC1,GFMC2}, both the central-type and tensor-type correlation functions employed in the tensor-optimized antisymmetrized molecular dynamics (TOAMD) \cite{TOAMD1,TOAMD2,TOAMD3,TOAMD4}, and the power-series-type correlation functions used in the tensor-optimized Fermi sphere (TOFS) method \cite{TOFS1,TOFS2,TOFS3}. The results calculated by GFMC and TOAMD are consistent with each other and the obtained energies of finite nuclei are reproduced very well, indicating the efficient descriptions of both the short-range correlation and the tensor correlation in finite nuclei. The TOFS method is developed to study the nuclear matter within a Fermi sphere approximation. The obtained EoS of the symmetric nuclear matter agrees with those of other benchmark calculations. Another popular approach to treat the $NN$ correlations is the renormalization of the $NN$ interaction by using unitary transformation, such as unitary correlation operator method (UCOM) \cite{UCOM1,UCOM2,UCOM3,UCOM4} and similarity renormalization group (SRG) transformation \cite{SRG1,SRG2,SRG3,SRG4,SRG5}. By introducing a unitary correlation operator multiplied to the trial wave function, the Hamiltonian of the system within UCOM can be transformed with the $NN$ correlations effectively treated. Then the eigenvalue can be obtained by solving the Schr\"{o}dinger equation with the transformed Hamiltonian and the trial wave function. It was shown that UCOM can effectively treat the short-range correlation in finite nuclei \cite{UCOM1,UCOM2,UCOM3,UCOM4}. Moreover, within the Hartree-Fock (HF) approximation for a Fermi sphere, UCOM has also been employed to successfully treat the short-range correlation in neutron matter \cite{Hu1,Hu2}. Similar to UCOM, the unitary operators are also introduced in SRG but a continuous unitary transformation with a flow parameter. The SRG is designed to soften the $NN$ interactions and decouple the low- and high-momentum scales in the system Hamiltonian. By calculating two-body densities in coordinate and momentum space, the evolution of the short-range correlation is investigated within the SRG transformations \cite{SRG4}. On the other hand, since the $NN$ correlations can excite the momenta of nucleon pairs into high-momentum (HM) region, resulting in HM components in finite nuclei and nuclear matter, two-particle two-hole (2p2h) excitations for nucleon pairs are recently developed to directly describe the $NN$ correlations \cite{2p2h1,2p2h2,2p2h3,2p2h4,2p2h5}. The two correlated nucleons produce a large transfer momentum with an opposite direction to each other, leading to a large relative momentum \cite{2p2h5}. The 2p2h excitations of such nucleon pairs with high momentum, which is called HM pairs hereafter, have been successfully employed in the tensor-optimized shell model (TOSM) \cite{TOSM2,TOSM3} and the high-momentum AMD (HMAMD) \cite{2p2h1,2p2h2,2p2h3,2p2h4} to study the properties of finite nuclei. By optimizing the 2p2h configurations without truncation of the particle states, the tensor correlation in finite nuclei can be effectively described and the total energies of the nuclei can be well reproduced.

According to the successful description of the $NN$ correlations in finite nuclei, some efficient approaches can also be used to study the properties of nuclear matter, such as the GFMC \cite{GFMC1}. There are several other popular many-body theories employed for nuclear matter, such as Brueckner-Hartree-Fock (BHF) \cite{BHF1,BHF2,BHF3}, Brueckner-Bethe-Goldstone (BBG) \cite{BHF2,BHF3,BHF4}, self-consistent Green's function (SCGF) \cite{SCGF1,SCGF2,SCGF3}, Fermi hypernetted chain/single-operator chain (FHNC/SOC) \cite{FHNC1,FHNC2,FHNC3}, auxiliary-field diffusion Monte Carlo (AFDMC) \cite{AFDMC1,AFDMC2,AFDMC3,AFDMC4}, and coupled cluster (CC) theory \cite{CC1,CC2,CC3}. With these famous theories, Baldo \emph{et al.} performed benchmark calculations with the family of widely used Argonne version (AVX) $NN$ potentials for both symmetric nuclear matter and neutron matter \cite{BHF3}. Recently, Piarulli \emph{et al.} also performed benchmark calculations for pure neutron matter with both the family of the AVX potentials and four chiral effective field theory potentials \cite{AFDMC4}. Very similar behaviors for the density dependence of the total energy per particle in symmetric nuclear matter and neutron matter are obtained. In detail, the energies at low densities are consistent with each other, while there are energy differences at high densities. Particularly, the saturation property of the symmetric nuclear matter shows a very close relation with the tensor force \cite{BHF3}. Besides of the bare two-body $NN$ force, the three-body force is also pointed out to be important for the exact saturation point of the symmetric nuclear matter \cite{AFDMC3}. Moreover, the discrepancies among above approaches at high densities are much larger for symmetric nuclear matter than those for neutron matter. This is mainly because of their different treatments of the stronger tensor correlation in symmetric nuclear matter. The intermediate- and long-range properties of the tensor force from the $NN$ interaction can induce many-body correlations, which are also important for nuclear matter studies.

In our recent works \cite{UCOMHM1,UCOMHM2}, the properties of nuclear matter are studied from bare $NN$ interactions by employing UCOM to treat the short-range correlation and introducing the 2p2h excitations of the HM pairs to describe the tensor correlation and spin-orbit coupling. This new developed method is named as UCOM+HM. By using AV4' potential which includes the strong short-range repulsion, the EoSs of both symmetric nuclear matter and neutron matter are systematically calculated \cite{UCOMHM1}. The obtained results are in good agreement with those of other popular theories, indicating the efficient treatment of the short-range correlation by UCOM. By employing AV6' and AV8' potentials including the tensor force, the effects of the short-range correlation, the tensor correlation, and the spin-orbit couplings on the EoS of neutron matter are investigated \cite{UCOMHM2}. In this paper, we will concentrate on the properties of the symmetric nuclear matter by including the tensor correlation. The tensor force is known to be much stronger for proton-neutron ($pn$) pairs than those for proton-proton ($pp$) and neutron-neutron ($nn$) pairs, hence the tensor correlation in symmetric nuclear matter are much more complicated than that in neutron matter. Usually, the momentum excitation mode for the correlated nucleons are employed to describe the $NN$ correlations in finite nuclei and nuclear matter. Here we will discuss more excitation modes for the HM pairs. Especially, since the tensor force is closely related to the two spins of the correlated nucleons, we will introduce the spin excitation mode to describe the strong tensor correlation in symmetric nuclear matter. This paper is organized as follows. In Sec. \ref{ucom}, the formalism of UCOM is presented. In Sec. \ref{2p2hex}, different 2p2h excitation modes of the HM pairs especially the spin excitation mode are introduced. The total wave function of the symmetric nuclear matter including the tail correction is also given. The calculated results are presented and discussed in Sec. \ref{results}. A summary is given in Sec. \ref{summary}.

\section{Framework}
\label{formula}

\subsection{Unitary correlation operator method (UCOM)}
\label{ucom}
The unitary correlation operator method (UCOM) is introduced to treat the short-range correlation in symmetric nuclear matter. Within the UCOM, the correlated wave function $\Psi$ can be obtained by $\Psi=C_r\Phi$, where $\Phi$ is the trial wave function and $C_r$ is the unitary correlation operator. The latter is defined as \cite{UCOM1,UCOM2}
\begin{align}\label{eqcr}
C_r=\exp\left(-i\sum_{i<j}g_{ij}\right)=\sum^A_{i<j}c_{r,ij},
\end{align}
where $g_{ij}$ is a pair-type Hermite generator and $c_{r,ij}$ is the operator for one pair in the $A$-body system. The specific form of the generator $g$ can be denoted as
\begin{align}\label{eqgij}
g=\frac{1}{2}\{p_rs(r)+s(r)p_r\},
\end{align}
where the operator $p_r$ is the relative momentum parallel to the relative coordinate of the correlated nucleons. The function $s(r)$ is the variation of the wave function for the relative motion at the distance $r$. In the calculations of UCOM, the function $s(r)$ is usually replaced by $R_+(r)$, which satisfies the following relations \cite{UCOM1,UCOM2}
\begin{align}\label{eqrpr}
\frac{dR_+(r)}{dr}&=\frac{s[R_+(r)]}{s(r)},\\
c^{\dag}_rrc_r&=R_+(r).
\end{align}
The function $R_+(r)$ represents the transformed distance from the original one $r$. It can decrease the amplitude of the wave function for the relative motion at short-range distances, as a result of the short-range correlation. The forms of $R_+(r)$ for even channel with positive parity and odd channel with negative parity are given as \cite{UCOM1,UCOM2}
\begin{align}\label{eqrprf}
R^{\textmd{even}}_+(r)&=r+\alpha\left(\frac{r}{\beta}\right)^{\gamma}\exp\left[-\exp\left(\frac{r}{\beta}\right)\right],\\
R^{\textmd{odd}}_+(r)&=r+\alpha\left(1-\exp\left[-\frac{r}{\gamma}\right]\right)\exp\left[-\exp\left(\frac{r}{\beta}\right)\right],
\end{align}
where the parameters $\alpha$, $\beta$, and $\gamma$ are variationally determined and their values for different channels in symmetric nuclear matter are listed in Table \ref{ta01}, which are the same to our previous work \cite{UCOMHM1}.

\begin{table}[thb]
\centering
\caption{Values of the parameters $\alpha$, $\beta$, and $\gamma$ in the function $R_+(r)$ of UCOM for symmetric nuclear matter.}
\label{ta01}
\begin{tabular}{cccc}
\hline\hline
\vspace{-5.3mm}
\kern13mm & \kern22mm & \kern22mm & \kern22mm  \\\hline\hline
       & $\alpha$   &  $\beta$   &  $\gamma$   \\\hline
$^1E$  & 1.36       &   0.98     &    0.33     \\
$^3E$  & 1.24       &   0.94     &    0.39     \\
$^1O$  & 1.50       &   1.26     &    0.87     \\
$^3O$  & 0.69       &   1.39     &    0.28     \\\hline\hline
\end{tabular}
\end{table}

With the determined function $R_+(r)$, the generator $g_{ij}$ and the operator $c_{r,ij}$ as well as the operator $C_{r}$ can be obtained. Then by using the transformation $\Psi=C_r\Phi$, the original Schr\"{o}dinger equation $H\Psi=E\Psi$ can be derived as $\tilde{H}\Phi=E\Phi$, where $\tilde{H}$ is the transformed Hamiltonian and can be denoted as \cite{UCOM1,UCOM2}
\begin{align}\label{eqhamil}
\tilde{H}&=C^{\dag}_rHC_r=C^{\dag}_rTC_r+C^{\dag}_rVC_r=\tilde{T}+\tilde{V},\\\label{eqhamil2}
\tilde{T}&=\sum^A_{i=1}t_i+\sum^A_{i<j}u_{ij},\\
\tilde{V}&=\sum^A_{i<j}\tilde{v}_{ij}.
\end{align}
In the transformed potential $\tilde{V}$, the radial part in $\tilde{v}_{ij}(r)$ is obtained by $v_{ij}(R_+(r))$, which is transformed from the original $v(r)$ at the relative distance $r$. The transformed kinetic energy operator $\tilde{T}$ contains two parts, the one-body part $t_i$ and the correlated two-body part $u_{ij}$. The latter comes from the short-range correlation between the correlated nucleons, which is closely related to both the momentum and angular momentum of the relative motion \cite{UCOM1,UCOM2}:
\begin{align}\label{equcomu}
u(r)=w(r)+\frac{1}{2}\left[p^2_r\frac{1}{2\mu_r(r)}+\frac{1}{2\mu_r(r)}p^2_r\right]+\frac{\boldsymbol{L}^2}{2\mu_{\Omega}(r)r^2},
\end{align}
where the forms of the functions $w(r)$, $\mu_r(r)$, and $\mu_{\Omega}(r)$ are respectively given using the nucleon mass $m$ as
\begin{align}\label{eqwpl}
w(r)&=\frac{\hbar^2}{m}\left(\frac{7}{4}\frac{R''^2_+(r)}{R'^4_+(r)}-\frac{1}{2}\frac{R'''_+(r)}{R'^3_+(r)}\right),\\
\frac{1}{2\mu_r(r)}&=\frac{1}{m}\left(\frac{1}{R'^2_+(r)}-1\right),\\
\frac{1}{2\mu_{\Omega}(r)}&=\frac{1}{m}\left(\frac{r^2}{R^2_+(r)}-1\right).
\end{align}

\subsection{2p2h excitation modes and total wave function}
\label{2p2hex}
The symmetric nuclear matter is described by the finite particle-number approach, where the periodical boundary condition is employed for the single-nucleon wave function as $\phi(\boldsymbol{r})=\phi(\boldsymbol{r}+L\hat{\boldsymbol{x}})$. Under this description, the infinite nuclear matter can be divided into identical cubic boxes with finite size $L$. The 0p0h state of the cubic box is defined by the Slater determinant as \cite{UCOMHM1,UCOMHM2}
\begin{align}\label{eq0p0h}
|\textmd{0p0h}\rangle&=\frac{1}{\sqrt{A!}}\textmd{det}\left\{\prod^A_{i=1}\phi_{\alpha_i}(\boldsymbol{r}_i)\right\},\\
\phi_{\alpha}(\boldsymbol{r})&=\frac{1}{\sqrt{L^3}}e^{i\boldsymbol{k}_{\alpha}\cdot\boldsymbol{r}}\chi^{\sigma}_{\alpha}\chi^{\tau}_{\alpha},\\
\langle\phi_{\alpha}|\phi_{\alpha'}\rangle&=\delta_{\alpha\alpha'},
\end{align}
where $A$ is the mass number in the cubic box. The single-nucleon wave function $\phi_{\alpha}(\boldsymbol{r})$ is described by a plane wave and the functions $\chi^{\sigma}_{\alpha}$ and $\chi^{\tau}_{\alpha}$ are the spin and isospin components, respectively. The index $\alpha$ represents the quantum number for the momentum, spin and isospin. Within the periodical boundary condition, the momenta of nucleons are discretized by the gap $\Delta k=\frac{2\pi}{L}$ and all the momenta are located at the grid points of the lattice in the momentum space. By using an integer vector $\boldsymbol{n}=(n_x, n_y, n_z)$, the momentum of each nucleon can be obtained by $\boldsymbol{k}=\frac{2\pi}{L}\boldsymbol{n}$. Constrained by the Fermi momentum $k_F$, there exist magic numbers of the total grid points in the momentum lattice, such as $N_g=1, 7, 19, 27, 33, 57, ...$. As a result of the spin and isospin symmetries in symmetric nuclear matter, the particle numbers in the cubic box can be $A=4N_g=4, 28, 76, 108, 132, 228, ...$.

The HM component in symmetric nuclear matter is described by introducing 2p2h excitations of nucleon pairs, which can be written as \cite{UCOMHM1,UCOMHM2}
\begin{align}\label{eq2p2h}
|\textmd{2p2h}\rangle=|mn;i^{-1}j^{-1}\rangle=a^{\dag}_ma^{\dag}_na_ia_j|\textmd{0p0h}\rangle,
\end{align}
where the indices $i$ and $j$ ($i,j=1,...,A$) represent hole states from lower magnitude of momenta, and the indices $m$ and $n$ ($m,n>A$) are particle states in the 2p2h configurations. The total momentum between the two holes and the two particles is conserved under the following condition
\begin{align}\label{eqmom1}
\boldsymbol{k}_i+\boldsymbol{k}_j&=\boldsymbol{k}_m+\boldsymbol{k}_n,\\\label{eqmom2}
\boldsymbol{k}_m&=\boldsymbol{k}_i+\boldsymbol{q},\\\label{eqmom3}
\boldsymbol{k}_n&=\boldsymbol{k}_j-\boldsymbol{q},
\end{align}
where the magnitudes of the above momenta satisfy
\begin{align}\label{eqmagt}
|\boldsymbol{k}_m|>k_F&,\quad|\boldsymbol{k}_n|>k_F,\\
|\boldsymbol{k}_i|<k_F&,\quad|\boldsymbol{k}_j|<k_F.
\end{align}
The quantity $\boldsymbol{q}$ is the transfer momentum between the two correlated nucleons, which links the two holes and two particles in the 2p2h configuration. The transfer momentum $\boldsymbol{q}=\frac{2\pi}{L}\boldsymbol{n}_q$ is also discretized with the integer momentum mode $\boldsymbol{n}_q=(n_{qx}, n_{qy}, n_{qz})$. If the transfer momentum $\boldsymbol{q}$ is large enough, the HM components in symmetric nuclear matter can be naturally induced by the 2p2h excitations. The maximum mode $n^{\textmd{max}}_q$ of the transfer momentum determines the total number of 2p2h configurations, which affects the basis space of present calculations.

\begin{figure}[thb]
\centering
  \includegraphics[width=1.0\linewidth,angle=0,clip=true]{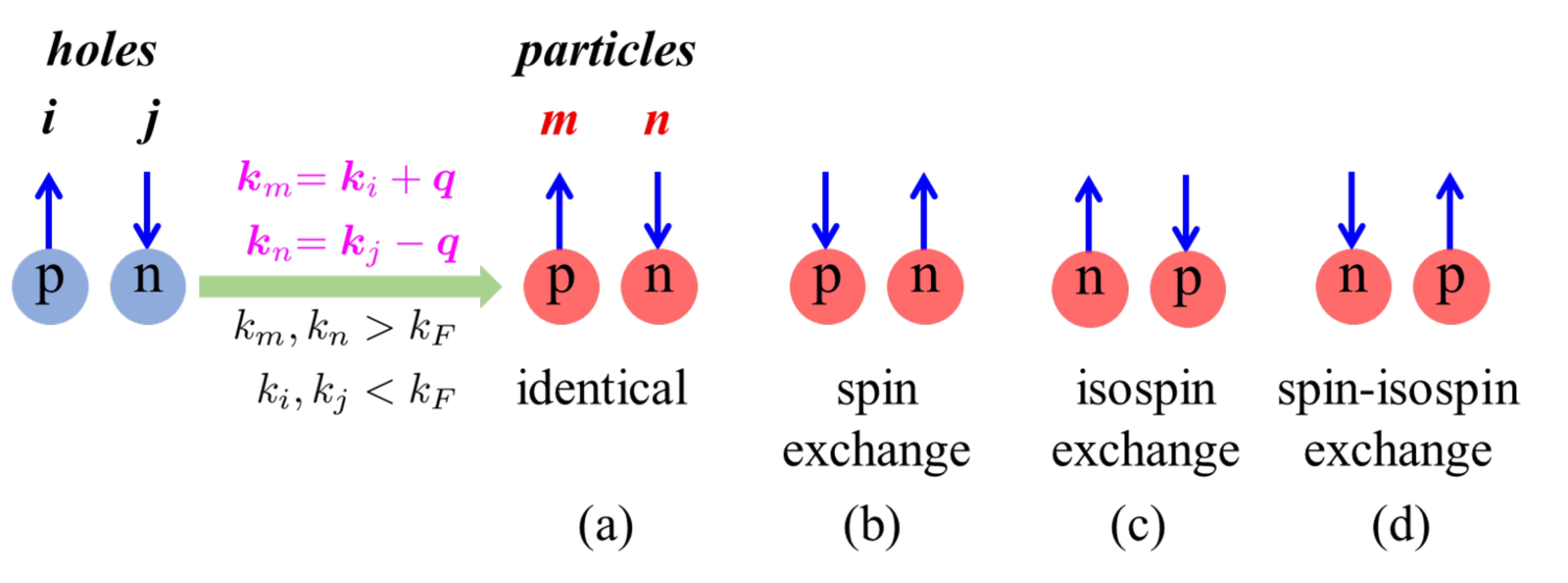}\\
  \caption{Sketch of momentum and spin/isospin exchange modes for 2p2h excitations.}
  \label{figexchange}
\end{figure}

The momentum excitations of two correlated nucleons are usually considered to describe the HM components in finite nuclei and nuclear matter. According to the $NN$ interaction, there are spin and isospin exchange terms in the central force. Therefore, the spin and isospin of the two correlated nucleons can be exchanged along with the above momentum excitations. As shown in Fig. \ref{figexchange}, by taking a $pn$ pair as an example, we show the momentum excitation and spin/isospin exchange modes for the 2p2h configurations. The blue up and down arrows represent the spin directions of the nucleons. As shown in Fig. \ref{figexchange}(a), only the momenta of the two correlated nucleons are excited into the HM region. Their spins and isospins are identical. In Fig. \ref{figexchange}(b) and (c), along with the momentum excitation, their spins/isospins are exchanged as well. In Fig. \ref{figexchange}(d), their spins and isospins are exchanged at the same time.

\begin{figure}[thb]
\centering
  \includegraphics[width=1.0\linewidth,angle=0,clip=true]{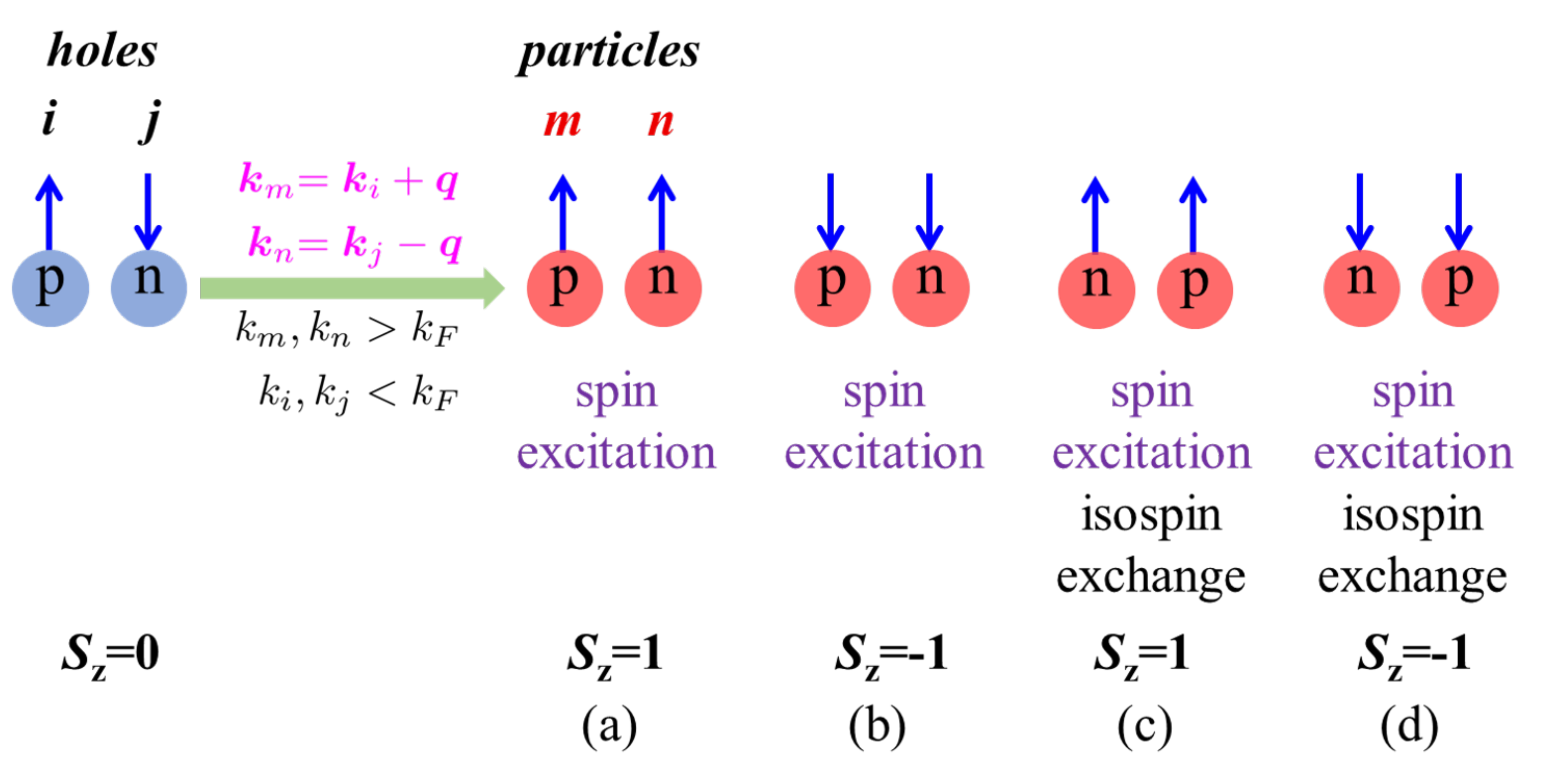}\\
  \caption{Sketch of spin excitations for $pn$ pairs with opposite directions of spins.}
  \label{figexcitation0}
\end{figure}

Besides of above momentum excitation, spin exchange, isospin exchange, and spin-isospin exchange modes for 2p2h configurations, we further consider spin excitation mode, which is important for treating the effect of the tensor force. As shown in Fig. \ref{figexcitation0}, we give the sketch of the spin excitation mode for a $pn$ pair with opposite spin direction. Hence the total spin of the nucleon pair before excitation is $S_z=0$. However, after the momentum excitation, there is possibility that the spin of the neutron can be excited from down to up. Hence the total spin of the HM pair is $S_z=1$, as shown in Fig. \ref{figexcitation0}(a). Similarly, as shown in Fig. \ref{figexcitation0}(b), the spin of the proton can be excited from up to down and the total spin is $S_z=-1$. In Fig. \ref{figexcitation0}(c) and (d), the isospins of the correlated two nucleons are exchanged along with their momentum and spin excitations. The total spins are $S_z=1$ and $S_z=-1$, respectively. For these excitations, the variation of the total spin after excitations is $\Delta S_z=1$.

\begin{figure}[thb]
\centering
  \includegraphics[width=1.0\linewidth,angle=0,clip=true]{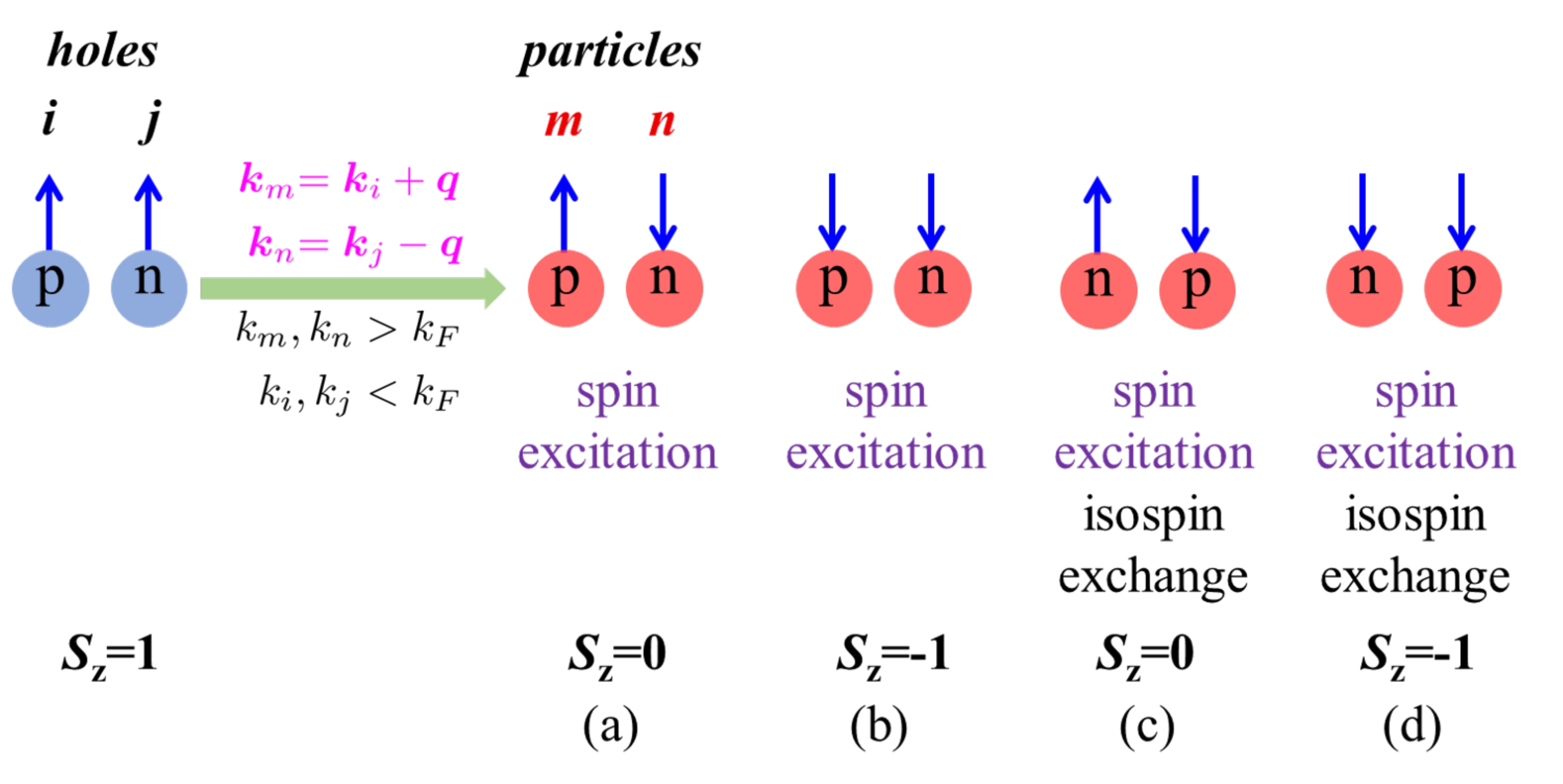}\\
  \caption{Sketch of spin excitations for $pn$ pairs with parallel directions of spins.}
  \label{figexcitation1}
\end{figure}

\begin{figure}[thb]
\centering
  \includegraphics[width=0.9\linewidth,angle=0,clip=true]{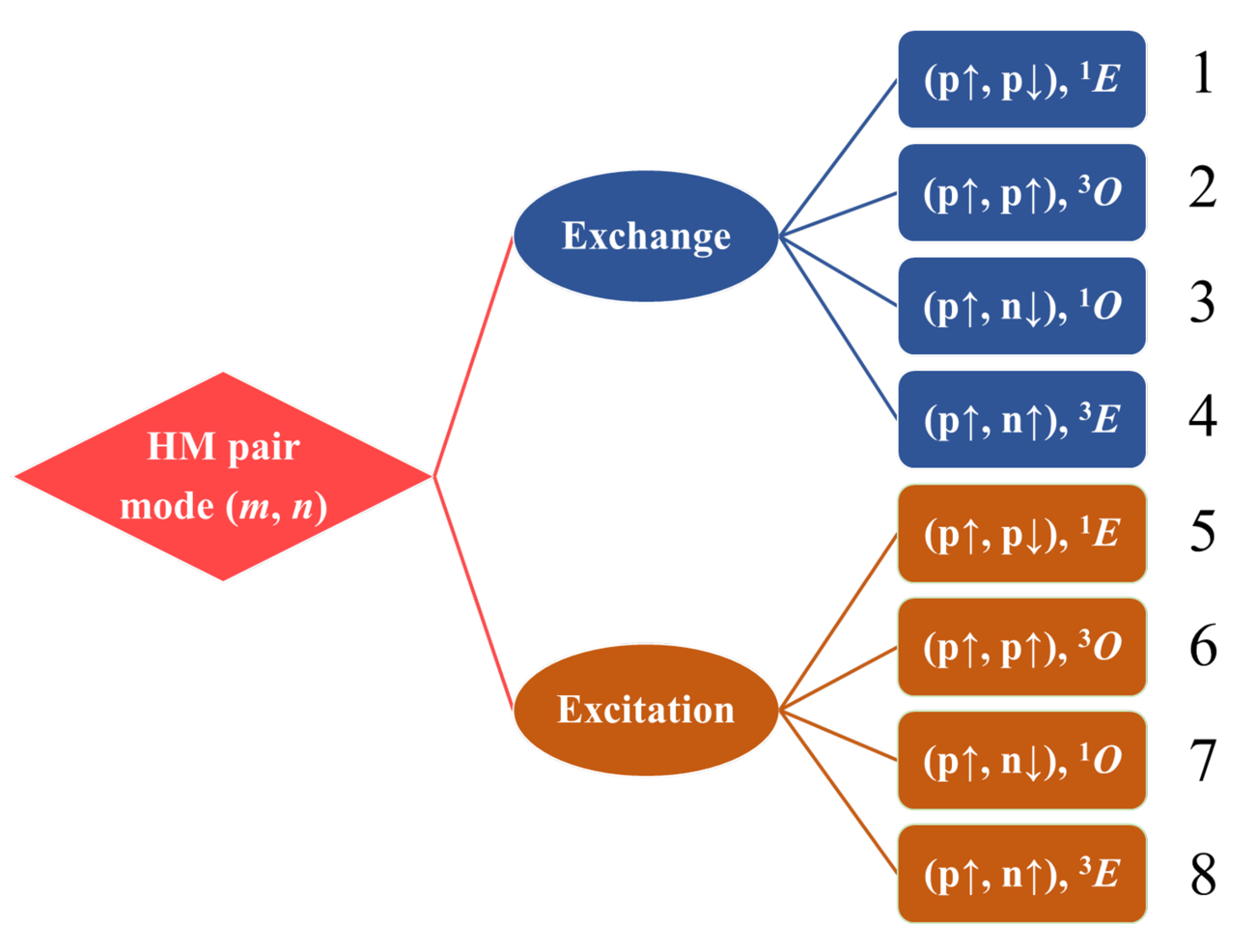}\\
  \caption{Different excitation modes and channels for the two particle state $(m,n)$ in 2p2h configurations.}
  \label{class}
\end{figure}

Shown in Fig. \ref{figexcitation1} is the sketch of the spin excitation mode for a $pn$ pair with parallel spin direction. Hence the total spin of the nucleon pair before excitation is $S_z=1$. If only one spin of the nucleon pair is excited along with the momentum excitation, as shown in Fig. \ref{figexcitation1}(a) and (c), the total spin of the nucleon pair is $S_z=0$ and the variation of the total spin is $\Delta S_z=1$. What's more, there is possibility that the two spin of the nucleon pair can both be excited at the same time, as shown in Fig. \ref{figexcitation1}(b) and (d). Hence the total spin is $S_z=-1$ and the variation of the total spin is $\Delta S_z=2$. Besides, as shown in Fig. \ref{figexcitation1}(c) and (d), the isospins of the nucleon pair are exchanged at the same time. Due to the isospin component, the excitation modes for $pn$ pairs are more complex than those for $pp$ and $nn$ pairs. In Fig. \ref{class}, based on the different excitation modes and nucleon-pair channels, we classify the HM pairs for the two particle state $(m,n)$ in 2p2h configurations into 8 modes. The four channels $^{1}E$, $^{3}O$, $^{1}O$, and $^{3}E$ for different types of nucleon pairs with spin/isospin exchange modes correspond to the mode indices 1--4. The channels with spin excitation mode are the mode indices 5--8. Considering the tensor force closely related to the spins of the correlated nucleons, the spin excitation mode is expected to effectively describe the strong tensor correlation in symmetric nuclear matter.

In addition, the isospin excitation mode for the 2p2h configurations was also considered. However, we found that the isospin excitations had none contributions to the total energy in present calculations. Besides, if the isospin excitation mode is included, the proton and neutron numbers will be changed, leading to different total nucleon numbers of the nuclear matter. In the future, we can further check the effects of the isospin excitation mode with the full Argonne $NN$ interaction which includes the isotensor (charge-breaking) components.

Moreover, since the tensor force is the interaction acting in intermediate- and long-distance regions, it is necessary for symmetric nuclear matter to consider the tail correction originating from the neighbouring boxes. The tail correction is usually estimated by extending the integrals for the two-body matrix elements from the box size $L$ to infinity \cite{AFDMC1}. As pointed out in Ref. \cite{AFDMC2}, the tail correction also can be included by tabulating both the two-body potential and the Jastrow factor within additional neighbouring boxes. The former one is similar to the integral extension for the two-body matrix elements, while the latter one includes the correlations coming from the neighbouring boxes. Hence, besides of the 2p2h configurations for the two correlated nucleons in the same box, there also exist correlations between two nucleons in different boxes. As a result of the tail correction from neighbouring boxes, the two correlated nucleons in different boxes also can form a HM pair with a 1p1h excitation in both boxes. In order to further describe the tensor correlation, it is necessary for symmetric nuclear matter to include the tail correction. Thus the momentum relations for the 2p2h configurations in Eqs. (\ref{eqmom2}) and (\ref{eqmom3}) can be rewritten as
\begin{align}\label{eqmom4}
\boldsymbol{k}_{ma}&=\boldsymbol{k}_{ia}+\boldsymbol{q},\\
\boldsymbol{k}_{nb}&=\boldsymbol{k}_{jb}-\boldsymbol{q}.
\end{align}
The total momentum of the associated 2p2h configuration is still conserved by the relation $\boldsymbol{k}_{ma}+\boldsymbol{k}_{nb}=\boldsymbol{k}_{ia}+\boldsymbol{k}_{jb}$, where $a$ and $b$ are the indices of involved boxes. For each associated 2p2h configuration, the particle $m$ and the hole $i$ are in the box $a$, and the other particle $n$ and the other hole $j$ are in the box $b$. Hence the correlated two nucleons are in the same box with $a=b$ and in different boxes with $a\neq b$. The spin/isospin exchange modes and the spin excitation modes discussed above also exist for the associated 2p2h configurations. As a result, all the cubic boxes for symmetric nuclear matter contain 0p0h, 1p1h, and 2p2h configurations. We have checked that in each single box, the 1p1h configurations have no couplings with 0p0h and 2p2h configurations. This is natural because the 1p1h excitations are induced by the interaction from the neighbouring boxes rather than the present box. However, the associated 2p2h excitations with two 1p1h configurations in different boxes can make contributions to the total energy of symmetric nuclear matter which will be discussed in the following section. This indicates the necessary inclusion of the tail correction for describing the tensor correlation in symmetric nuclear matter.

By including the 2p2h configurations, the total wave function of symmetric nuclear matter can be written as
\begin{align}\label{eqttwf}
\Phi=C_0|\textmd{0p0h}\rangle+\sum^{N_{\textmd{2p2h}}}_{p=1}C_p|\textmd{2p2h},p,b\rangle,
\end{align}
where $N_{\textmd{2p2h}}$ is the total number of 2p2h configurations and $\{C_p\}$ are the configuration amplitudes which can be variationally determined. The index $p$ represents each configuration with $p=0$ corresponding to the 0p0h state. The index $b$ ($1\leq b\leq N_{\textmd{b}}$) represents the involved box and $N_{\textmd{b}}$ is the total box number which can be variationally determined as well.

\begin{figure}[thb]
\centering
  \includegraphics[width=0.8\linewidth,angle=0,clip=true]{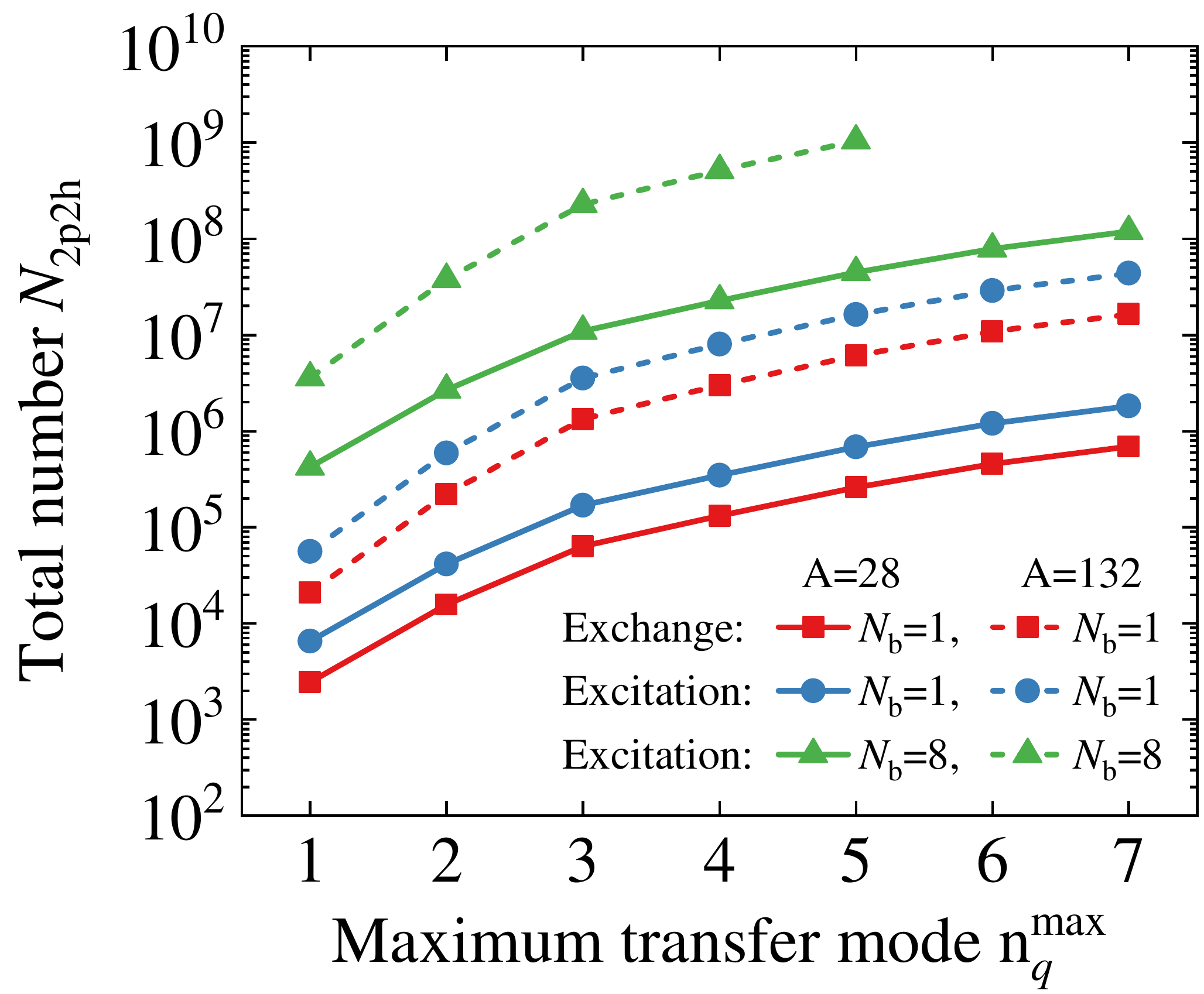}\\
  \caption{Total number of 2p2h configurations $N_{\textmd{2p2h}}$ varied with the maximum transfer mode $n^{\textmd{max}}_q$ for symmetric nuclear matter with different excitation modes and involved boxes.}
  \label{N2p2h}
\end{figure}

In Fig. \ref{N2p2h}, by taking the mass numbers $A=28$ and $A=132$ as examples, we plot the total number of the 2p2h configurations $N_{\textmd{2p2h}}$ varied with the maximum transfer mode $n^{\textmd{max}}_q$ for different excitation modes and different numbers of involved boxes. It is clearly shown in Fig. \ref{N2p2h} that the total configuration number $N_{\textmd{2p2h}}$ is generally exponentially increased with the maximum transfer mode $n^{\textmd{max}}_q$. Besides, for the central box with $N_{\textmd{b}}=1$, compared with the total configuration number without spin excitation mode (red square lines), the inclusion of the spin excitation mode (blue circle lines) can increase the number by several times for the same transfer momentum mode. When the tail correction is considered with more boxes in the calculations, such as for $N_{\textmd{b}}=8$, the total configuration number will be further increased by several orders of magnitude and come up to $10^8$. Hence the basis space in present calculations will be very huge. Besides, as pointed out by GFMC \cite{GFMC1}, AFDMC \cite{AFDMC2,AFDMC3}, CC \cite{CC2}, and UCOM+HM in our previous works \cite{UCOMHM1,UCOMHM2}, the cubic box with mass number $A=132$ can best simulate the properties of the HF infinite nuclear matter. However, our previous work has shown that by using UCOM to treat the short-range correlation, the relative errors of both the kinetic and potential energies are reduced, especially for the potential energies \cite{UCOMHM1,UCOMHM2}. Hence, the energies with different magic mass numbers agree with each other without much accuracy loss. Considering the available computing capability, we will take the mass number $A=28$ to study the properties of symmetric nuclear matter in the following calculations.

By using the basis states giving in Eq. (\ref{eqttwf}), we can calculate the matrix elements of the transformed Hamiltonian $\tilde{H}$. Physically, the UCOM transformation can induce two-body correlations. Then by introducing 2p2h configurations into the total wave function, there are up to 4p4h correlations involved in present UCOM+HM approach. Within the power method \cite{Power1,Power2,Power3}, we can solve the eigenvalue problem for the Hamiltonian matrix with $N_{\textmd{2p2h}}+1$ dimensions. Then the configuration amplitudes $\{C_p\}$ and the total number of boxes $N_\textmd{b}$ involved in Eq. (\ref{eqttwf}) can be variationally determined by minimizing the total energy of symmetric nuclear matter. With the obtained values, we can further calculate the EoSs of symmetric nuclear matter with different interactions and the contributions of each Hamiltonian component to the total energy.

\section{Results and discussion}
\label{results}

\begin{figure*}[thb]
\centering
  \includegraphics[width=0.8\linewidth,angle=0,clip=true]{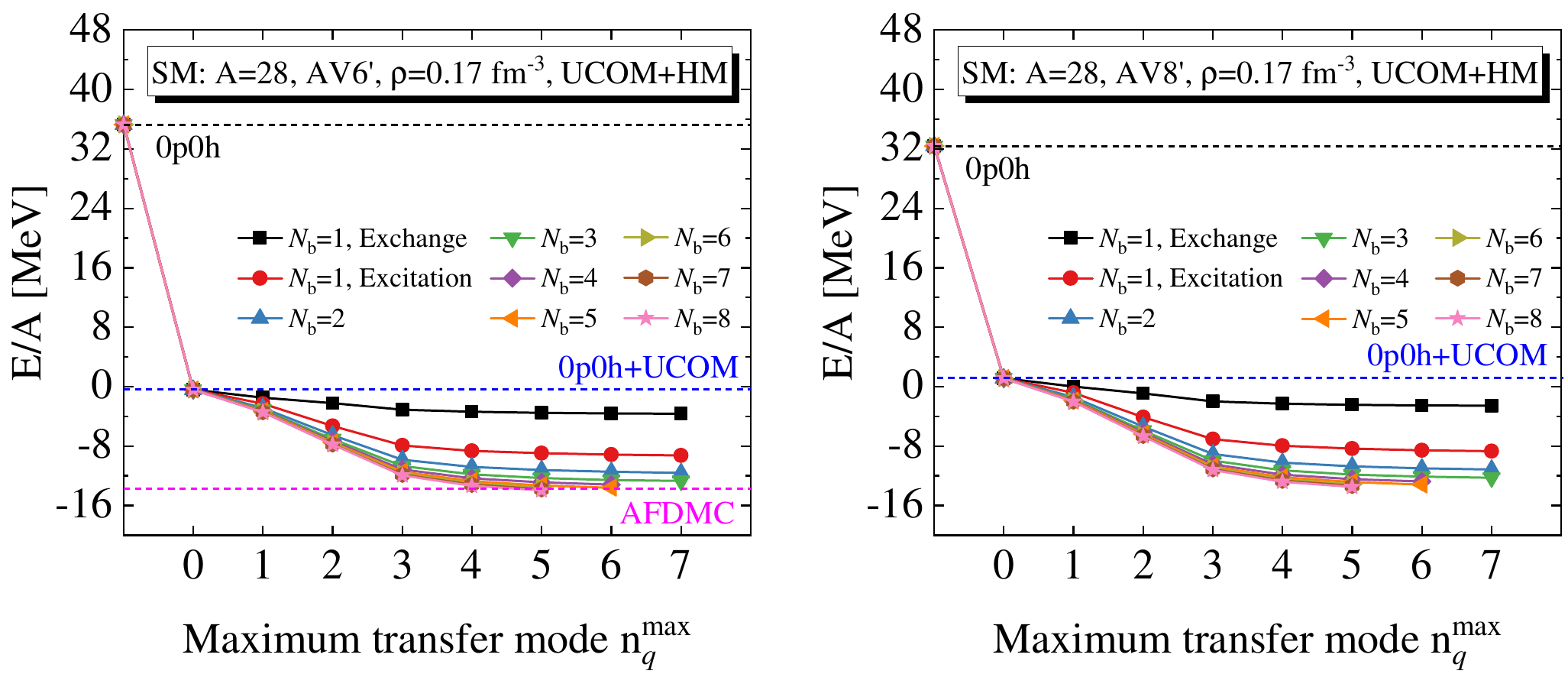}\\
  \caption{Convergence of the total energy per particle for symmetric nuclear matter with the maximum transfer mode $n^{\textmd{max}}_q$ and the number of involved boxes $N_\textmd{b}$ under AV6' (left) and AV8' (right) potentials.}
  \label{AV68cvg}
\end{figure*}

\begin{figure*}[thb]
\centering
  \includegraphics[width=0.8\linewidth,angle=0,clip=true]{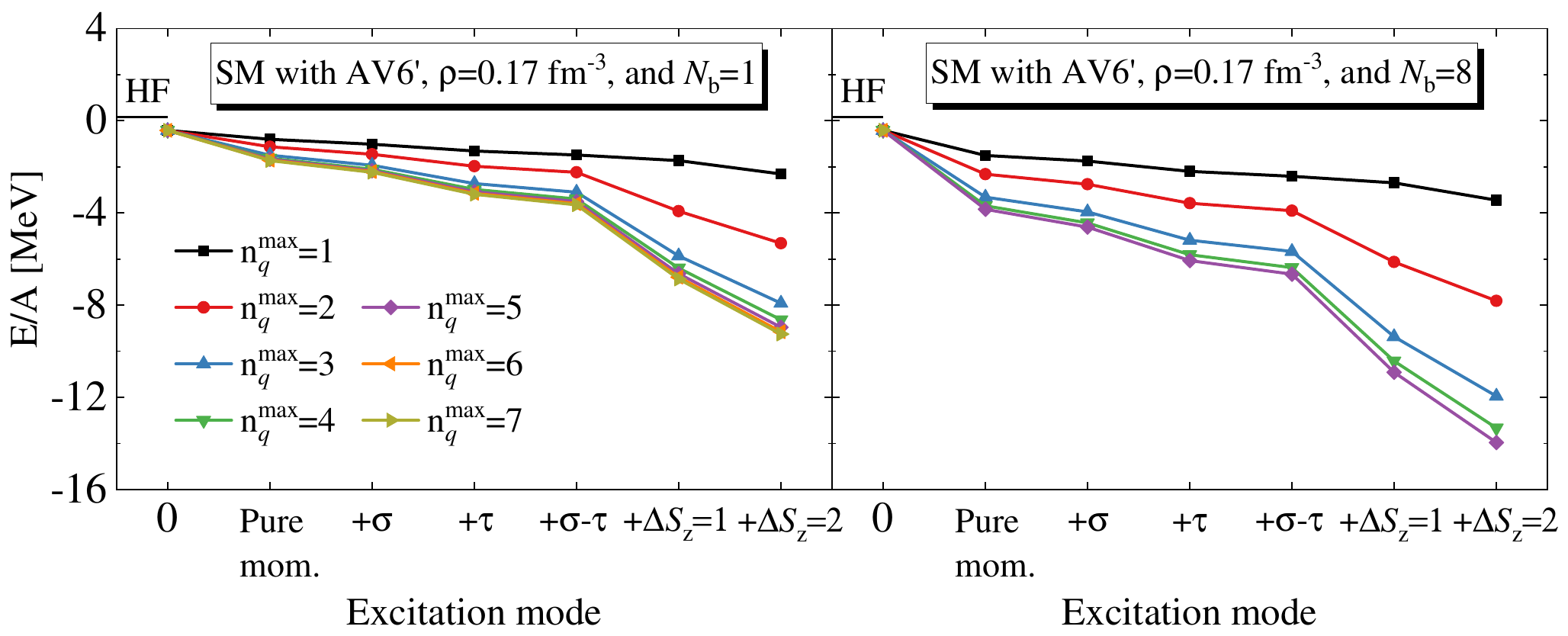}\\
  \caption{Contributions of different excitation modes to the total energy per particle at the normal density $\rho=0.17\ \textmd{fm}^{-3}$ in symmetric nuclear matter with AV6' potential by taking the box numbers $N_\textmd{b}=1$ (left) and $N_\textmd{b}=8$ (right) as examples.}
  \label{exmode}
\end{figure*}

In Fig. \ref{AV68cvg}, we plot the energy convergence of symmetric nuclear matter with the maximum transfer mode $n^{\textmd{max}}_q$ and the number of involved boxes $N_\textmd{b}$ under AV6' and AV8' potentials, respectively. The results are calculated at the normal density $\rho=0.17\ \textmd{fm}^{-3}$ with mass number $A=28$. It can be directly seen from Fig. \ref{AV68cvg} that for both AV6' and AV8' potentials, the total energy converges with the maximum transfer mode $n^{\textmd{max}}_q$ increasing. This is reasonable because the excitation probability will decrease with the increasing transfer momentum. What's more, the total energy will also converge with the increasing box number $N_\textmd{b}$. This indicates that the tail correction from the neighbouring boxes will be saturated due to the finite range of the $NN$ interactions.

In each panel of Fig. \ref{AV68cvg}, we also give the results calculated with 0p0h (black dash lines) and 0p0h+UCOM (blue dash lines) wave functions at the normal density. With 0p0h wave function for symmetric nuclear matter, the total energy per particle at the normal density is about 35.2 MeV with AV6' potential and 32.4 MeV with AV8' potential. By using UCOM to treat the short-range correlation, there exists a large energy fall from 0p0h wave function to 0p0h+UCOM wave function, as shown from the black dash line to the blue dash line in each panel of Fig. \ref{AV68cvg}. This is the result of the effective treatment of the short-range correlation and the corresponding total energy per particle is decreased to -0.4 MeV with AV6' potential and 1.1 MeV with AV8' potential, respectively. Then we introduce the nucleon pair excitations from both the central box and the neighbouring boxes to describe the tensor correlation in symmetric nuclear matter. The final converged energy per particle is around -13.8 MeV with AV6' potential and -13.3 MeV with AV8' potential, respectively. The obtained value is very close to the calculation with AFDMC, as shown by the magenta dash line in the left panel.

\begin{figure*}[thb]
\centering
  \includegraphics[width=0.8\linewidth,angle=0,clip=true]{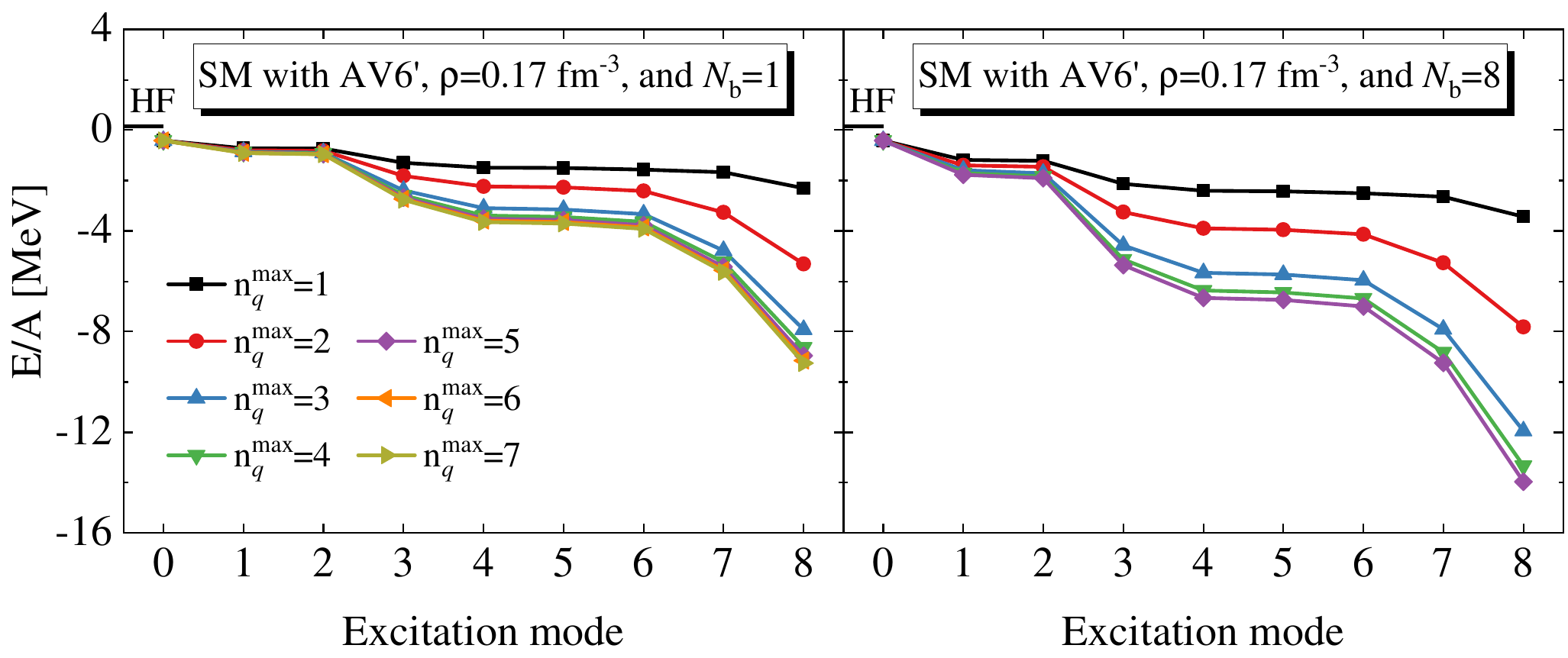}\\
  \caption{Contributions of different excitation modes to the total energy per particle at the normal density $\rho=0.17\ \textmd{fm}^{-3}$ in symmetric nuclear matter with AV6' potential by taking the box numbers $N_\textmd{b}=1$ (left) and $N_\textmd{b}=8$ (right) as examples.}
  \label{exmode2}
\end{figure*}

\begin{figure*}[thb]
\centering
  \includegraphics[width=0.7\linewidth,angle=0,clip=true]{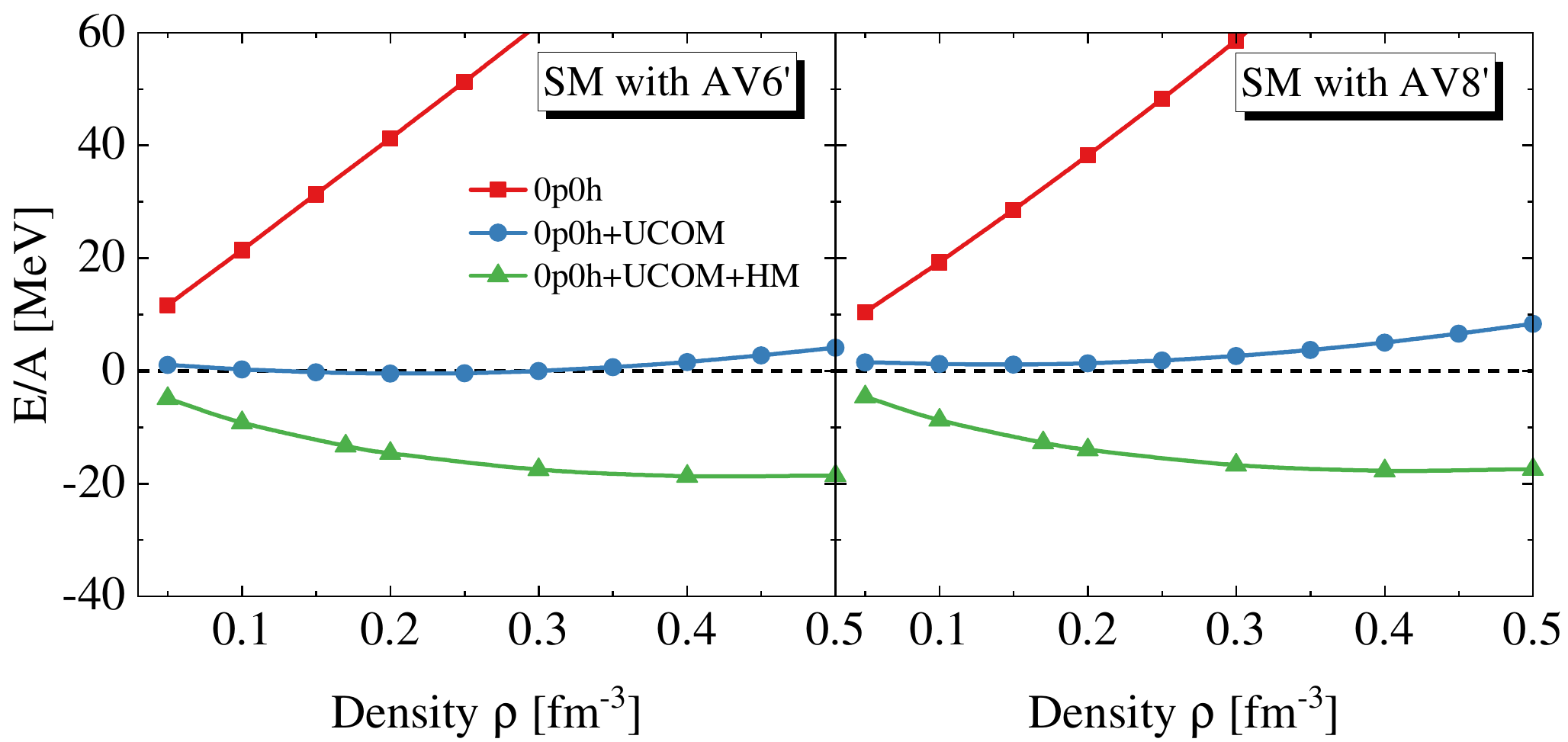}\\
  \caption{Comparison of the density dependence of the total energy per particle in symmetric nuclear matter calculated with different wave functions under AV6' (left) and AV8' (right) potentials.}
  \label{smwf}
\end{figure*}

Besides, in each panel of Fig. \ref{AV68cvg}, the black square lines represent the results only in the central box without considering the spin excitation mode of HM pairs. The converged energy is about -3.6 MeV with AV6' potential and -2.6 MeV with AV8' potential. For the results of the red circle lines, the spin excitation mode is included in the total wave function but still only in the central box. The corresponding converged energy is decreased as -9.3 MeV with AV6' potential and -8.7 MeV with AV8' potential. Hence the spin excitation mode can make significant contributions to the total energy of symmetric nuclear matter. Since the short-range correlation has been treated by UCOM, the contributions of the spin excitation mode mainly come from the effective description of the tensor correlation. This indicates the particular importance of the spin excitation mode for the tensor correlation in symmetric nuclear matter. For the results with box number $N_\textmd{b}$ from 2 to 8, the spin excitation mode is also included in all the calculations and the total energy with AV6' potential is finally converged around the result calculated with the AFDMC method. By comparing the results with $N_\textmd{b}=1$ and with $N_\textmd{b}=8$, we can find that the converged energy gains from the tail correction are about 4.2 MeV. Hence, besides of the spin excitations, the tail correction from the neighbouring boxes also has remarkable contributions to the total energy and to the tensor correlation in symmetric nuclear matter.

By taking the box numbers $N_\textmd{b}=1$ and $N_\textmd{b}=8$ as examples, we study the effects of different excitation modes for the HM pairs. The contributions of each mode to the total energy per particle at the normal density $\rho=0.17\ \textmd{fm}^{-3}$ with AV6' potential are shown in Fig. \ref{exmode}. The energies with excitation mode ``0" correspond to the 0p0h+UCOM calculations. The other excitation modes along the horizontal axis are pure momentum excitation without spin-isospin exchange, spin exchange, isospin exchange, spin-isospin exchange, spin excitation with $\Delta S_z=1$, and spin excitation with $\Delta S_z=2$. It can be seen from Fig. \ref{exmode} that by successively adding the excitation modes into the total wave function, the total energy per particle will be decreased, indicating the individual contributions from these excitation modes. Besides, with the maximum transfer momentum mode $n^{\textmd{max}}_q$ increasing, the contributions will converge. This is also because of the decreasing possibility of the 2p2h excitations with larger transfer momentum. Moreover, it is obvious in Fig. \ref{exmode} that there are relatively large energy gains for the spin excitation modes with $\Delta S_z=1$ and $\Delta S_z=2$, which originate from the effective description of the tensor correlation by the spin excitation mode. In addition, for the mode with $\Delta S_z=2$, both the spins of the correlated nucleons should be excited to the opposite direction. Thus the excitation possibility could be much smaller than that of the mode with $\Delta S_z=1$. However, the former has approximately the same contribution to the total energy per particle, indicating the significance of the spin excitation mode with $\Delta S_z=2$ for the tensor correlation. Besides, when the involved box increases from $N_\textmd{b}=1$ to $N_\textmd{b}=8$, there is a rise for the contribution from each excitation mode. This is due to the tail correction from the neighbouring boxes. Besides, the contributions from the pure momentum excitation and spin excitation modes are generally larger than those of the spin exchange, isospin exchange, and spin-isospin exchange modes.

\begin{figure*}[thb]
\centering
  \includegraphics[width=0.7\linewidth,angle=0,clip=true]{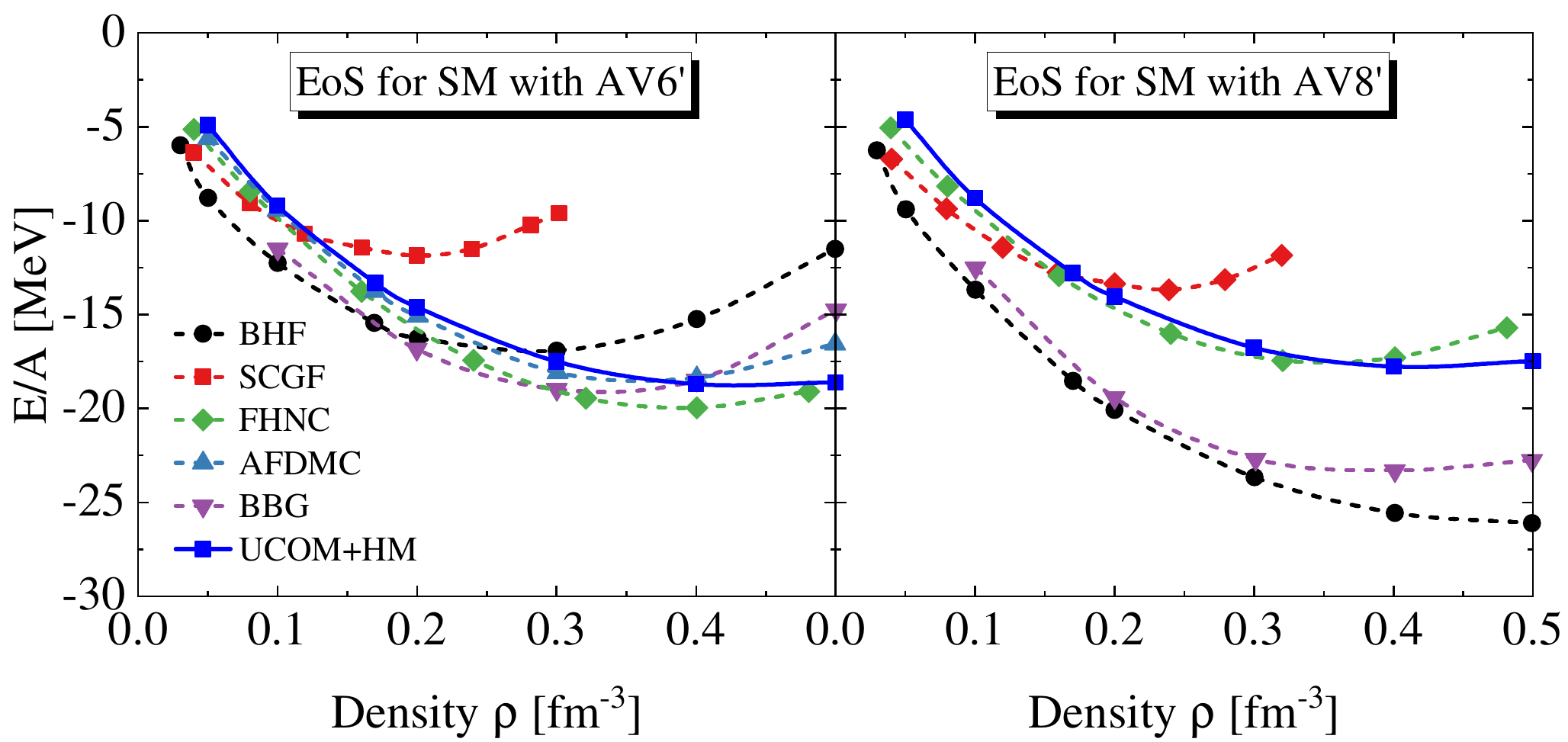}\\
  \caption{Equations of state of symmetric nuclear matter calculated under AV6' (left) and AV8' (right) potentials.}
  \label{smeos}
\end{figure*}

\begin{figure*}[thb]
\centering
  \includegraphics[width=0.7\linewidth,angle=0,clip=true]{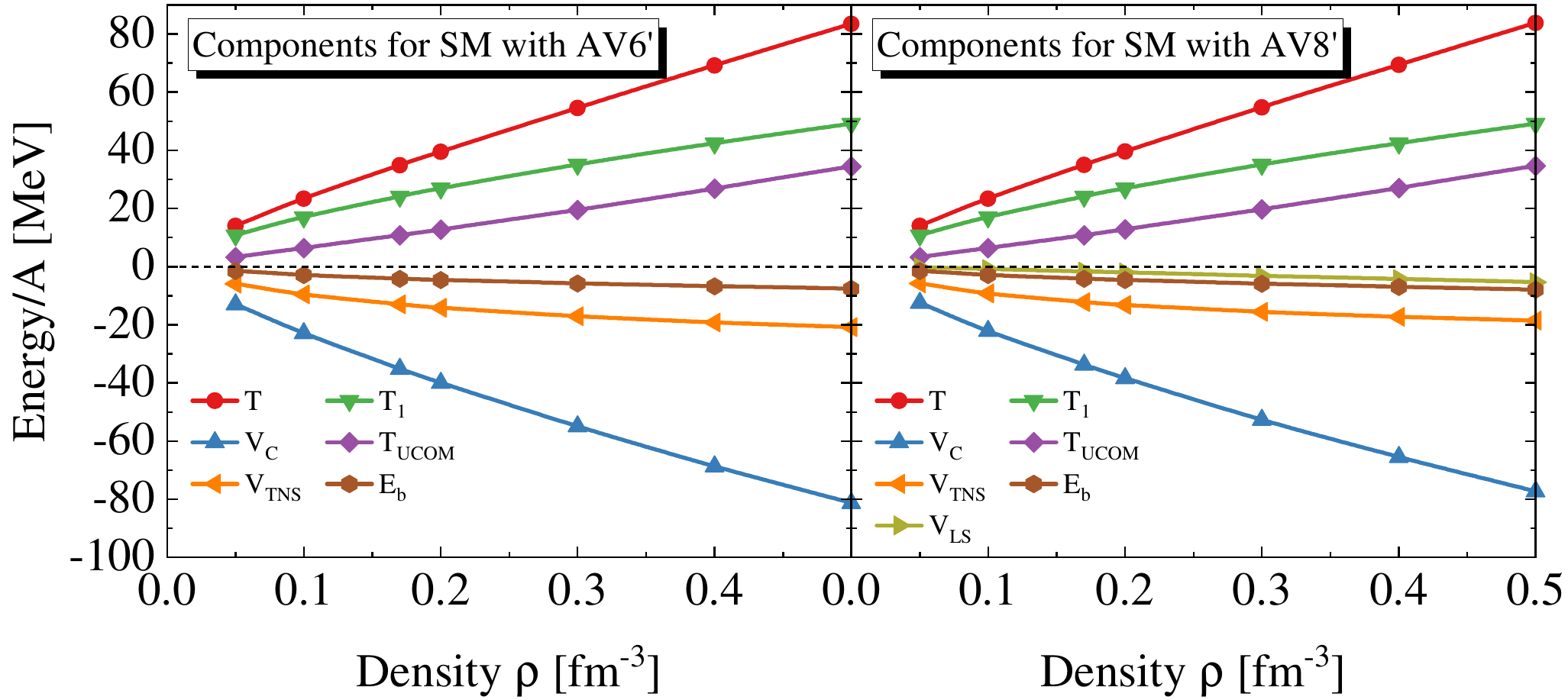}\\
  \caption{Energies of the Hamiltonian components for symmetric nuclear matter calculated under AV6' (left) and AV8' (right) potentials. $T$ is the total kinetic energy, $T_1$ is the summation of the one-body kinetic energy $t_i$, and $T_{\textmd{UCOM}}=T-T_1$ is the two-body kinetic energy originating from the short-range correlation in UCOM. $V_\textmd{C}$, $V_{\textmd{TNS}}$, and $V_{\textmd{LS}}$ are the central, tensor, and spin-orbit coupling parts of potential energy, respectively. The quantity $E_{\textmd{b}}=E(N_\textmd{b}=8)-E(N_\textmd{b}=1)$ represents the contribution of the tail correction from the neighbouring boxes.}
  \label{smcmpnt}
\end{figure*}

In order to study the tensor correlations for different $NN$ pairs, we plot in Fig. \ref{exmode2} the contributions of the different HM pairs to the total energy in symmetric nuclear matter. The energies with excitation mode ``0" are still calculated with the 0p0h+UCOM wave function. The indices ``1--4" correspond to the four channels $^{1}E$, $^{3}O$, $^{1}O$, and $^{3}E$ from the exchange modes, while ``5--8" are the four channels within spin excitation modes, as shown in Fig. \ref{class}. The former two channels $^{1}E$ and $^{3}O$ are for $pp$/$nn$ pairs, while the latter two $^{1}O$ and $^{3}E$ are for $pn$ pairs. It is clearly shown in Fig. \ref{exmode2} that the $pn$ pairs have significant contributions to the total energy while the contributions from the $pp$/$nn$ pairs are very minor. This is mainly because of the stronger tensor correlation for $pn$ pairs than for $pp$/$nn$ pairs. Besides, the energy gains from the $pn$ pairs with spin excitations 7 and 8 are obviously larger than those from exchange modes 3 and 4. This also indicates the significance of the spin excitation modes for the tensor correlation in symmetric nuclear matter.

In Fig. \ref{smwf} we show the comparison of the total energy per particle calculated with 0p0h, 0p0h+UCOM, and (0p0h+)UCOM+HM wave functions under AV6' and AV8' potentials, respectively. It can be seen that for each potential, the energies with 0p0h wave function are all positive and much stiffer. This is because of the very strong short-range repulsion from the central force. After the short-range correlation is treated by UCOM, there are very large energy falls at all densities with 0p0h+UCOM wave function. When the HM pairs are introduced, the total energies are further decreased, indicating the effective description of the tensor correlation in symmetric nuclear matter.

\begin{table*}[thb]
\centering
\caption{Values of the Hamiltonian components as well as the total energy per particle for symmetric nuclear matter at some densities with AV6' and AV8' potentials. The units of the energies are $\textmd{MeV}/A$.}\label{ta02}
\begin{tabular}{ccccccccccc}
\hline\hline
\vspace{-4mm}
\kern13mm  &\kern13mm  &\kern13mm  &\kern13mm   &\kern13mm  &\kern4mm  &\kern13mm  &\kern13mm  &\kern13mm  &\kern13mm  &\kern13mm    \\
$\rho$    & \multicolumn{4}{c}{AV6'}       &          &  \multicolumn{5}{c}{AV8'}\\\cline{2-5}\cline{7-11}
$(\textmd{fm}^{-3})$    &   E    &   T    & $\textmd{V}_{\textmd{C}}$   &  $\textmd{V}_{\textmd{TNS}}$ & &   E    &   T    & $\textmd{V}_{\textmd{C}}$    &  $\textmd{V}_{\textmd{TNS}}$ &  $\textmd{V}_{\textmd{LS}}$ \\\hline
 0.05  & -4.92  & 13.99 & -12.96 & -5.95  & & -4.63  & 13.97 & -12.52 & -5.80  & -0.28 \\
 0.10  & -9.20  & 23.34 & -22.92 & -9.62  & & -8.78  & 23.37 & -22.13 & -9.20  & -0.82 \\
 0.17  & -13.33 & 34.83 & -35.13 & -13.03 & & -12.81 & 34.92 & -33.82 & -12.22 & -1.69 \\
 0.20  & -14.62 & 39.49 & -39.96 & -14.15 & & -14.05 & 39.61 & -38.42 & -13.18 & -2.06 \\
 0.30  & -17.51 & 54.54 & -54.98 & -17.07 & & -16.76 & 54.73 & -52.64 & -15.61 & -3.24 \\
 0.40  & -18.70 & 69.16 & -68.66 & -19.20 & & -17.78 & 69.39 & -65.50 & -17.33 & -4.34 \\
 0.50  & -18.62 & 83.49 & -81.26 & -20.85 & & -17.47 & 83.76 & -77.26 & -18.63 & -5.34 \\
\hline\hline
\end{tabular}
\end{table*}

Shown in Fig. \ref{smeos} are the EoSs of symmetric nuclear matter for AV6' and AV8' potentials calculated with present UCOM+HM as well as several other many-body theories. The results of other approaches are taken from previous benchmark calculations \cite{BHF3}. From the left panel for AV6' potential, we can see that the obtained results in present work are consistent with those of other theories. In detail, the energies among different approaches agree with each other at low densities, while there exist differences at high densities. It should be noted that the present results are calculated with the mass number $A=28$ instead of $A=132$ where the latter can give the best simulation for the properties of the HF infinite nuclear matter. As systematically studied in our previous work \cite{UCOMHM1}, there is a common feature that the mass number $A=28$ will give a little lower total energy than $A=132$ at high densities, which is also obtained in the two panels of Fig. \ref{smeos}. From the right panel for AV8' potential, we can see that though there are agreements among different theories at small density region, the energy difference increases as the nuclear matter density increases. Besides, it is well known that there is no saturation point for symmetric nuclear matter with AV4' potential until very large density \cite{BHF3}. However, for AV6' and AV8' potentials including the tensor force, there exist saturation behaviors around the density $\rho=0.4\ \textmd{fm}^{-3}$. In order to show the saturation property more clearly, the numerical results for AV6' and AV8' potentials are listed in Table \ref{ta02}. This manifests the very close relationship between the tensor correlation and the saturation properties of symmetric nuclear matter. Besides of the tensor force, the three-body force is also pointed out to has impacts on the saturation properties. In the future, we can further include the three-body force into the calculations.

By decomposing the Hamiltonian into different components, we obtain their separate contributions to the total energy of symmetric nuclear matter. The results with AV6' and AV8' potentials are shown in the two panels of Fig. \ref{smcmpnt}, respectively. $T$ is the total kinetic energy, $T_1$ is the summation of the one-body kinetic energy $t_i$ given in Eq. (\ref{eqhamil2}), and $T_{\textmd{UCOM}}=T-T_1$ is the two-body kinetic energy originating from the short-range correlation in UCOM. $V_\textmd{C}$, $V_{\textmd{TNS}}$, and $V_{\textmd{LS}}$ are the central, tensor, and spin-orbit coupling parts of potential energy, respectively. The quantity $E_{\textmd{b}}=E(N_\textmd{b}=8)-E(N_\textmd{b}=1)$ represents the contribution of the tail correction from the neighbouring boxes. It can be seen from Fig. \ref{smcmpnt} that the total kinetic energy $T$ and the central potential energy $V_\textmd{C}$ make main contributions to the total energy, while the contributions of the tensor and spin-orbit coupling parts are relatively small. However, since the former two contributions are in opposite signs and canceled, the total energy is very close to the tensor part. Besides, the two-body kinetic energy $T_{\textmd{UCOM}}$ is comparable to the one-body one $T_1$, indicating the significance of the short-range correlation. As a result, the total kinetic energy $T$ increases with the density $\rho$ much faster than the one-body kinetic energy $T_1$ which generally goes as $\rho^{2/3}$. In addition, the contribution of the tail correction shown by the brown lines is even larger than that of the spin-orbit coupling part. Hence it is necessary to include the associated 2p2h configurations to further describe the tail correction and tensor correlation originating from the neighbouring boxes.

\section{Summary}
\label{summary}
We calculate the equation of state (EoS) of symmetric nuclear matter by using the bare AV6' and AV8' nucleon-nucleon ($NN$) interactions. The symmetric nuclear matter is described within the finite particle-number approach by using the periodical boundary condition for the single-nucleon wave function. Hence the infinite nuclear matter is divided into identical cubic boxes.

The unitary correlation operator method (UCOM) is used to treat the short-range correlation and the excitations of correlated nucleon pairs with high momentum (HM pairs) are employed to describe the tensor correlation in symmetric nuclear matter. Different excitation modes are analyzed for the HM pair excitations, including momentum excitation without spin-isospin exchange, spin exchange, isospin exchange, spin-isospin exchange, and spin excitation. The spin of the HM pairs are changeable by the spin excitation and the variations of the spin of the correlated pairs can be $\Delta S_z=1$ and $\Delta S_z=2$. Besides, in order to include the tail correction originating from the neighbouring boxes, the associated 2p2h configurations with two 1p1h excitations from different boxes are also considered in the total wave function of symmetric nuclear matter.

By calculating the total energy per particle of symmetric nuclear matter at the normal density $\rho=0.17\ \textmd{fm}^{-3}$, we first confirm the energy convergence with the maximum transfer momentum mode and the number of the involved boxes. The contributions of the short-range correlation and the HM pairs as well as the tail correction to the total energy are obtained. Then we calculate the contributions of different excitation modes to the total energy. The pure momentum excitation and spin excitation modes are found to have larger contributions than those of other excitation modes. In particular, the spin excitation mode is found to be very important for describing the tensor correlation in symmetric nuclear matter. The effects of the individual HM pairs on the total energy are also studied. Due to the strong tensor correlation for $pn$ pairs, the contributions from $pn$ pairs are found to be more significant than those from $pp$ and $nn$ pairs. Moreover, the $pn$ pairs from spin excitation modes have larger contributions than those from exchange modes.

With 0p0h, 0p0h+UCOM, 0p0h+UCOM+HM wave functions, we also compare the density dependence of the total energy per particle in symmetric nuclear matter for AV6' and AV8' potentials. The energies with 0p0h wave function are very stiff throughout the densities. However, the energies with 0p0h+UCOM wave function are decreased a lot due to the treatment of the short-range correlation by UCOM. When the HM pairs are used, the energies in symmetric nuclear matter will be further decreased because of the description of the tensor correlation.

The EoS of symmetric nuclear matter calculated under AV6' potential in present work is consistent with those of other many-body theories. In detail, the energies at low densities agree with each other, while there are differences among different approaches at high densities. For AV8' potential, the energies per particle with different approaches agree with each other at small density region, while the difference increases as the nuclear matter density increases. Moreover, the saturation properties of symmetric nuclear matter are confirmed to show close relation to the tensor correlation.

The contributions of each Hamiltonian component to the total energy of symmetric nuclear matter are obtained as well. The total kinetic energy and the central potential energy make the main contributions, but with opposite signs and canceled. As a result, the total energy is very close to the tensor potential energy. The correlated two-body kinetic energy is found to be comparable to the one-body kinetic energy, indicating the significance of the short-range correlation. The contribution of the tail correction to the total energy is remarkable and larger than the spin-orbit coupling potential energy. This indicates the necessity of the tail correction to further describe the tensor correlation originating from the neighbouring boxes.

\section*{ACKNOWLEDGEMENTS}
This work is supported by the National Natural Science Foundation of China (Grants No. 12205105 and No. 11822503), by the Fundamental Research Funds for the Central Universities (Nanjing University), by JSPS KAKENHI Grants No. JP18K03660 and No. JP22K03643, and by the startup funding of South China University of Technology.

\end{document}